\documentclass[aps,prl,twocolumn,nofootinbib,superscriptaddress,floatfix]{revtex4-2}
\usepackage{graphicx}
\usepackage{subfigure}
\usepackage{bbm}
\usepackage{amsmath}
\usepackage{amssymb}
\usepackage{amsfonts}
\usepackage{mathrsfs}
\usepackage{theorem}
\usepackage{bm}
\usepackage{url}
\usepackage[T1]{fontenc}
\usepackage{csquotes}
\MakeOuterQuote{"}

\usepackage{algorithm}
\usepackage{algorithmicx}
\usepackage{algpseudocode}

\usepackage{dcolumn}
\usepackage{color}
\usepackage[colorlinks,citecolor=blue]{hyperref}

\newcommand{\tr}{\operatorname{tr}}

\newcommand{\rmi}{\mathrm{i}}
\newcommand{\rme}{\mathrm{e}}

\newcommand{\rmP}{\mathrm{P}}
\newcommand{\rmR}{\mathrm{R}}
\newcommand{\rmT}{\mathrm{T}}

\newcommand{\caH}{\mathcal{H}}

\newcommand{\Rnum}[1]{\mathrm{\uppercase\expandafter{\romannumeral #1\relax}}}

\newcommand{\be}{\begin{equation}}
\newcommand{\ee}{\end{equation}}
\newcommand{\ba}{\begin{align}}
\newcommand{\ea}{\end{align}}

\def\<{\langle}  
\def\>{\rangle}  

\def\eqref#1{\textup{(\ref{#1})}}  
\newcommand{\eref}[1]{Eq.~\textup{(\ref{#1})}}

\newcommand{\esref}[1]{Eqs.~\textup{(\ref{#1})}}

\newcommand{\cref}[1]{Conjecture~\ref{#1}}
\newcommand{\Cref}[1]{Conjecture~\ref{#1}}

\newcommand{\rcite}[1]{Ref.~\cite{#1}}

\begin{document}

\title{Experimental Masking of Real Quantum States}

\author{Rui-Qi Zhang}
\thanks{These authors contributed equally to this work.}
\affiliation{Key Laboratory of Quantum Information,University of Science and Technology of China, CAS, Hefei 230026, P. R. China}
\affiliation{CAS Center For Excellence in Quantum Information and Quantum Physics}
\author{Zhibo Hou}
\thanks{These authors contributed equally to this work.}
\affiliation{Key Laboratory of Quantum Information,University of Science and Technology of China, CAS, Hefei 230026, P. R. China}
\affiliation{CAS Center For Excellence in Quantum Information and Quantum Physics}

\author{Zihao Li}
\thanks{These authors contributed equally to this work.}
\affiliation{State Key Laboratory of Surface Physics and Department of Physics, Fudan University, Shanghai 200433, China}
\affiliation{Institute for Nanoelectronic Devices and Quantum Computing, Fudan University, Shanghai 200433, China}
\affiliation{Center for Field Theory and Particle Physics, Fudan University, Shanghai 200433, China}

\author{Huangjun Zhu}
\email{zhuhuangjun@fudan.edu.cn}
\affiliation{State Key Laboratory of Surface Physics and Department of Physics, Fudan University, Shanghai 200433, China}
\affiliation{Institute for Nanoelectronic Devices and Quantum Computing, Fudan University, Shanghai 200433, China}
\affiliation{Center for Field Theory and Particle Physics, Fudan University, Shanghai 200433, China}
\author{Guo-Yong Xiang}
\email{gyxiang@ustc.edu.cn}
\affiliation{Key Laboratory of Quantum Information,University of Science and Technology of China, CAS, Hefei 230026, P. R. China}
\affiliation{CAS Center For Excellence in Quantum Information and Quantum Physics}
\author{Chuan-Feng Li}
\affiliation{Key Laboratory of Quantum Information,University of Science and Technology of China, CAS, Hefei 230026, P. R. China}
\affiliation{CAS Center For Excellence in Quantum Information and Quantum Physics}
\author{Guang-Can Guo}
\affiliation{Key Laboratory of Quantum Information,University of Science and Technology of China, CAS, Hefei 230026, P. R. China}
\affiliation{CAS Center For Excellence in Quantum Information and Quantum Physics}

\begin{abstract}
Masking of quantum information is a way of hiding information in correlations 
such that no information is accessible to any local observer. 
Although the set of all quantum states as a whole cannot be masked into bipartite 
correlations according to the no-masking theorem, the set of real states is 
maskable and is a maximal maskable set. 
In this work, we experimentally realize a masking protocol of the real ququart 
by virtue of a photonic quantum walk. 
Our experiment clearly demonstrates
that quantum information of the real ququart can be completely hidden in 
bipartite correlations of two-qubit hybrid entangled states, 
which are encoded in two different degrees of freedom 
of a single photon. 
The hidden information is not accessible from each 
qubit alone, but can be faithfully retrieved with a fidelity of 
about $99\%$ from correlation measurements. By contrast, 
any superset of the set of real density matrices cannot be 
masked. 
\end{abstract}
\date{\today}
\maketitle

\emph{Introduction.}---Hiding information in correlations is a useful idea that 
plays crucial roles in a wide spectrum of subjects.  Although classical 
information can  be completely hidden in  quantum correlations of a bipartite 
system,  quantum information of an arbitrary state cannot be 
completely hidden in a similar way according to the no-hiding theorem 
\cite{NoHide} and  no-masking theorem \cite{NoMask}. These no-go theorems 
offer valuable insights on the power and limitation of quantum information 
processing, which are complementary to the famous no-cloning theorem 
\cite{Wootters_1982,Dieks_1982,Lamas_Linares_2002} and no-broadcasting 
theorem \cite{Barnum_1996}. Meanwhile, they are of intrinsic interest to many 
active research areas, including quantum  communication, quantum secret 
sharing \cite{Hillery_1999,Cleve_1999,DiVincenzo_2002},  
information scrambling, and the black-hole information paradox 
\cite{Page_1993,Hayden_2007,Sekino_2008,Liu_2018}.

Notwithstanding the limitations set by the no-hiding and no-masking theorems 
\cite{NoMask}, it is possible to hide and mask quantum information in certain 
restricted sets of quantum states 
\cite{LiJiang_2019,Liang_2019,Ding_2020,Liang_2020,Du_2020,ZHU_Mask_Real} (see also  Ref.~\cite{LiWang_2018} for information masking in the multipartite scenario). 
Notably, very recently \rcite{ZHU_Mask_Real} showed that
the no-hiding and no-masking theorems break down in real quantum mechanics, 
in sharp contrast with complex quantum mechanics. Information 
about real density matrices (with respect to the computational basis) can be 
completely hidden in bipartite correlations. Moreover, the set of real states 
is a maximal maskable set in the sense that it is not contained in any other 
maskable set. These results give a twist to the problem of information masking, which is of intrinsic interest to the resource
theory of imaginarity \cite{Hickey_2018,wu2020operational} and foundational 
studies on quantum mechanics \cite{Wootters_1986,Chiribella_2011,hardy2001quantum} in addition to the research areas mentioned above.  
Despite these theoretical progresses, experimental works on 
quantum information masking are quite rare 
\cite{exp_masking1_2020,exp_masking2_2020} 
(see Refs.~\cite{exp_nohiding_2011,exp_nohiding_2019} on the test of the no-hiding 
theorem).  Moreover, all masking protocols implemented so far are restricted 
to qubit systems, which is a severe limitation.


In this work, we experimentally realize for the first time the masking protocol 
of the real ququart proposed in \rcite{ZHU_Mask_Real}  using a photonic system. 
To achieve this goal, we devise a four-step photonic quantum walk to realize
the desired masking isometry, which turns any pure state of the real ququart 
into a two-qubit hybrid entangled state.
Our experimental results clearly demonstrate 
that quantum  information of the real ququart can be completely hidden in 
bipartite correlations, and no information can be retrieved from each subsystem 
alone. In addition, the encoded quantum information can be faithfully retrieved 
from correlation measurements. By contrast, the output state associated with any 
input state that is not real cannot be maximally entangled, so  partial 
information has to leak to each subsystem. Moreover, the concurrence of the output state is determined by the robustness of imaginarity of 
the input state, which is of key interest to  the resource
theory of imaginarity \cite{Hickey_2018,wu2020operational}. 
Here, we encode two-qubit hybrid entangled states in
two different degrees of freedom (DoFs) of a single photon 
\cite{single_photon_2qubit1, single_photon_2qubit2}, 
but it can be generalized to two-photon two-qubit states by simply 
cascading the protocol of quantum state fission 
 \cite{quantum_fission} after our photonic quantum walk.

\emph{Masking of quantum information.}---A set $\mathscr{S}$ of quantum states 
on the Hilbert space $\mathcal{H}$ 
of dimension $d$ is maskable if there exists an 
isometry $M$ from $\mathcal{H}$ to $\mathcal{H}_A\otimes\mathcal{H}_B$, such 
that both $\tau_A = {\rm tr}_B(M\rho M^{\dagger})$ 
and $\tau_B = {\rm tr}_A(M\rho M^{\dagger})$ are independent of 
$\rho$ for all $\rho \in \mathscr{S}$ \cite{NoMask}. In this case, no information 
about the original state $\rho$  can be retrieved from subsystem  $A$ or $B$ alone, 
and all information spreads over 
the bipartite correlations between $A$ and $B$. Let $\mathscr{D}(\mathcal{H})$ 
and $\mathscr{P}(\mathcal{H})$ 
be the sets of all density matrices and all  pure states (rank-1 projectors) 
on $\mathcal{H}$;  let $\mathscr{D}^\rmR(\mathcal{H})$ and 
$\mathscr{P}^\rmR(\mathcal{H})$ be the sets of all real density matrices and 
all real pure states with respect to the computational basis 
$\{|j\rangle\}_{j=0}^{d-1}$ of $\mathcal{H}$. 
The no-masking theorem  states that $\mathscr{D}(\mathcal{H})$ 
and $\mathscr{P}(\mathcal{H})$ are not maskable \cite{NoMask}. However, 
this theorem does not apply to  restricted sets of quantum states in general. 
Notably, the set $\mathscr{D}^\rmR(\mathcal{H})$  is maskable and is actually 
a maximal maskable subset of $\mathscr{D}(\mathcal{H})$ \cite{ZHU_Mask_Real}. 
Here a  maskable set is maximal if  it is not a proper subset of any other 
maskable set. Such maximal maskable sets are particularly appealing because 
they reflect the potential and  limitation of quantum information masking as well 
as the distinction between quantum information processing and classical 
information processing. 

Since any mixed state is a convex mixture of pure states, masking of 
$\mathscr{D}^\rmR(\mathcal{H})$ is equivalent to the masking of 
$\mathscr{P}^\rmR(\mathcal{H})$. According to \rcite{ZHU_Mask_Real}, 
the set $\mathscr{P}^\rmR(\mathcal{H})$ can be masked by virtue of a 
set $\{U_j\}_{j=1}^{d-1}$ of $d-1$ unitary Hurwitz-Radon (HR) matrices 
acting on $\caH_A$ \cite{hurwitz1922komposition,HR}, which is characterized 
by the equation
\begin{equation}\label{eq:HR}
U_j U_k+U_k U_j=-2\delta_{jk}\mathbbm{1}_A.
\end{equation}
Let $|\Phi\rangle=\sum_{r=0}^{m-1} |rr\rangle$ be the canonical maximally entangled 
state in $\caH_A\otimes \caH_B$, assuming  that both $\caH_A$ and $\caH_B$ have 
dimension $m$. Let $U_0:=\openone_A$ and define 
\begin{equation}
|\Phi_j\rangle:=(U_j\otimes \openone_B)|\Phi\rangle,\quad j=0,1,\ldots, d-1. 
\end{equation}
Then the isometry $M$ from $\caH$ to $\mathcal{H}_A\otimes\mathcal{H}_B$ defined 
by the map $j\mapsto |\Phi_j\rangle$ is a masker for 
$\mathscr{P}^\rmR(\mathcal{H})$ and $\mathscr{D}^\rmR(\mathcal{H})$. Note that, 
for any normalized vector $\bm{c} := (c_0, c_1, ..., c_{d-1})$, the ket
$|\psi(\bm{c})\rangle = \sum_{j=0}^{d-1}{c_j|j\rangle}$ is mapped to 
$[U(\bm{c})\otimes \openone_B]|\Phi\rangle$ under the masker $M$, where $U({\boldsymbol c}) := \sum_j{c_jU_j}$. 
Moreover, 
$U(\bm{c})$ is a unitary operator for any normalized real vector $\bm{c}$ 
thanks to the characteristic of the HR matrices presented in \eref{eq:HR}.

\emph{Masking of the real ququart.}---In the case of $d = 4$, to mask 
$\mathscr{D}^\rmR(\mathcal{H})$, we need a set of three HR matrices. A 
simple example can be constructed from Pauli matrices: 
$\{U_j\}_{j=1}^{3}:=\{\rmi\sigma_z, \rmi\sigma_x, \rmi\sigma_y\}$, which can also be written as $\{\rmi Z, \rmi X, \rmi Y\}$.
With this 
choice, we can define an isometry 
$M: \mathcal{H}\mapsto \mathcal{H}_A\otimes\mathcal{H}_B$ as follows:
\begin{equation}\label{eq:masker}
	M|j\rangle = -\rmi|\Phi_j\rangle := -\rmi(U_j\otimes\mathbbm{1}_B)|\Phi\rangle,\quad j = 0,1,2,3,
\end{equation}
where $|\Phi\rangle := (|00\rangle + |11\rangle)/\sqrt{2}$ 
and $U_0 := \mathbbm{1}_A$. Here the choice of the phase factor $-\rmi$ is not essential but 
is convenient for the following discussion.
Under this isometry, the computational basis of $\mathcal{H}$ is mapped to the 
two-qubit magic basis \cite{,Magic_basis}; accordingly, all real pure states are 
mapped to maximally entangled states. So the isometry $M$ is a masker for 
$\mathscr{P}^\rmR(\mathcal{H})$ 
and $\mathscr{D}^\rmR(\mathcal{H})$ \cite{ZHU_Mask_Real}. By contrast, the 
image of any state in 
$\mathscr{D}(\mathcal{H})\setminus\mathscr{P}^\rmR(\mathcal{H})$ is not 
maximally entangled. Moreover, when $\rho$ is pure, the concurrence of the output state
$M\rho M^\dag$ is determined by the robustness of imaginarity of 
$\rho$ \cite{ZHU_Mask_Real},
\begin{equation}\label{eq:ConRI}
C(M\rho M^\dag)=\sqrt{1-\mathscr{I}_{\rmR}^2(\rho)},
\end{equation}
where $\mathscr{I}_R(\rho)=\|\rho-\rho^\rmT\|_1/2=\sqrt{1-\tr(\rho\rho^\rmT)}$ \cite{wu2020operational}. 

\begin{figure}[htbp]
	\center{\includegraphics[scale=0.23]{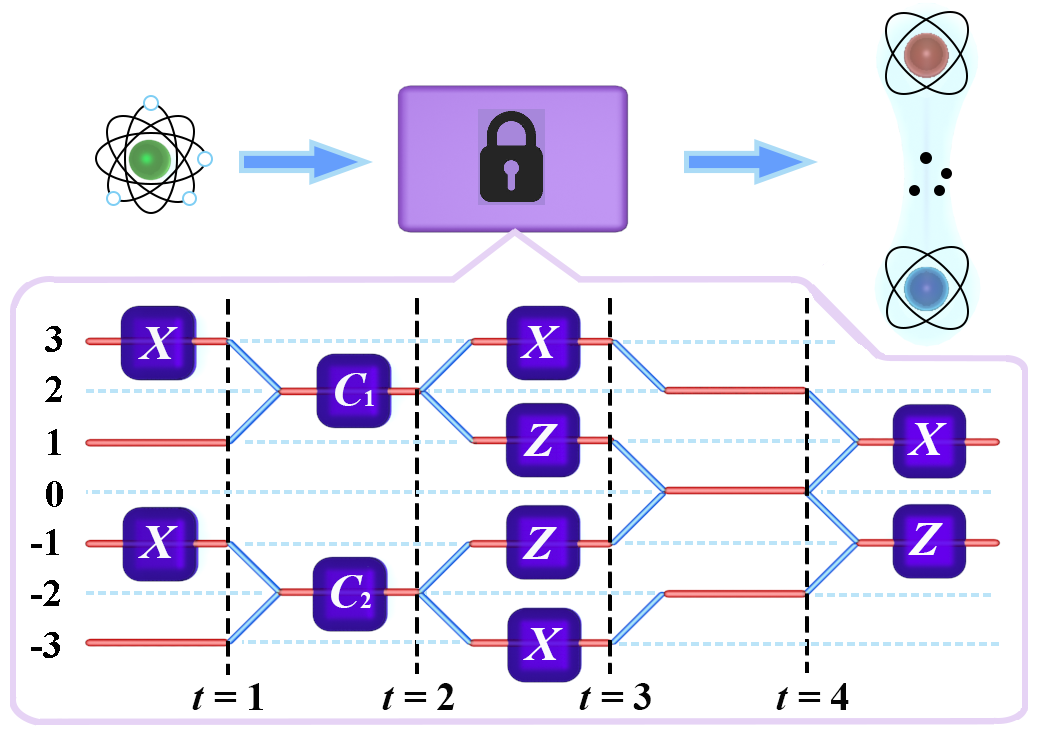}}
	\caption{\label{fig:quantum_walk} 
	Schematic diagram of quantum information masking (upper plot) and realization 
	of the masker $M$ of the real ququart based on a quantum walk (lower plot). 
The coin operators featuring in the figure 
are given by $X = \left(\begin{smallmatrix}0 & 1 \\ 1 & 0\end{smallmatrix}\right)$, 
	$Z = \left(\begin{smallmatrix}1 & 0 \\ 0 & -1\end{smallmatrix}\right)$, 
	${C}_1 = \frac{1}{\sqrt{2}}\left(\begin{smallmatrix}\rmi & 1 \\ -\rmi & 1 \end{smallmatrix}\right)$, and
	${C}_2 = \frac{1}{\sqrt{2}}\left(\begin{smallmatrix}1 & \rmi \\ -1 & \rmi \end{smallmatrix}\right)$. 
	}	
\end{figure}

\begin{figure*}[htbp]
	\center{\includegraphics[scale=0.225]{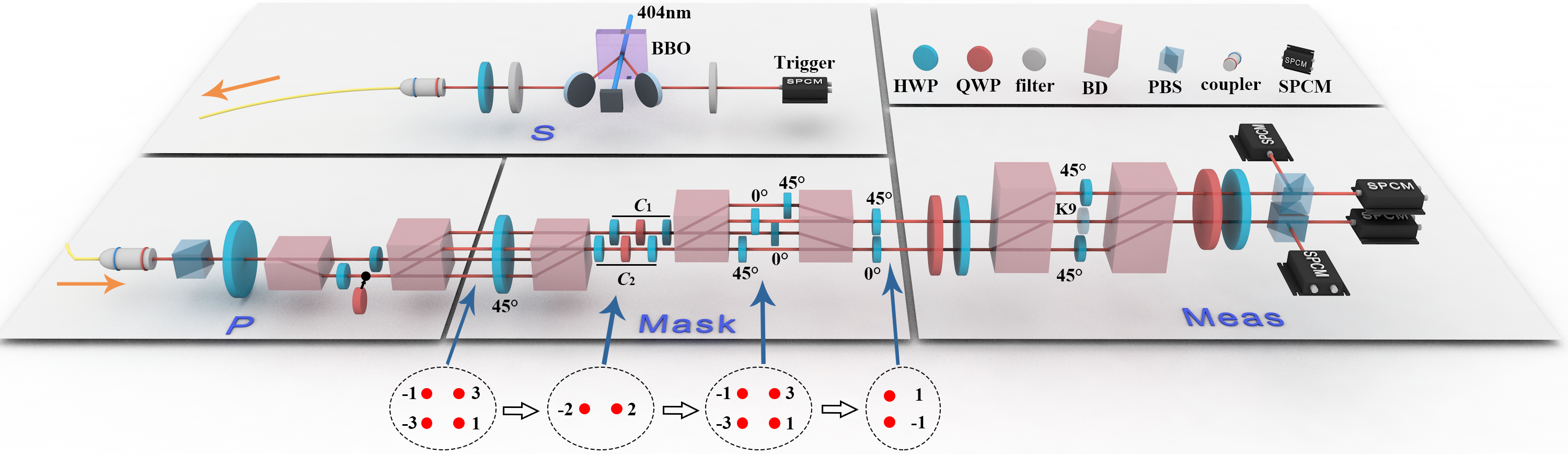}}
	\caption{\label{fig:Experimental setup}
		Experimental setup. The heralded single-photon source (labeled by S) 
		on top is realized by spontaneous parametric down-conversion in a 
		type-I $\beta$-barium-borate (BBO) crystal. The rest setup
		consists of  three modules: state-preparation module (labeled by P), 
		masking module (labeled by Mask), and measurement module 
		(labeled by Meas). The masking module implements the masker defined in 
		\eref{eq:masker} according to the quantum-walk scheme illustrated in 
		Fig.~\ref{fig:quantum_walk}. The photon's spatial modes labeled 
		at the bottom correspond to the walker's positions defined in 
		Fig.~\ref{fig:quantum_walk}. Two  combinations of wave plates of the  form 
		HWP-QWP-HWP in the masking module realize the coin operators $C_1$ and $C_2$ 
		defined in the caption of  Fig.~\ref{fig:quantum_walk}. HWP: half-wave plate; 
		QWP: quarter-wave plate; BD: beam-displacer; PBS: polarizing 
		beam-splitter; SPCM: single-photon counting module; K9: K9 glass plate.
	}
\end{figure*}

To realize the masker $M$ of the real ququart,  here we design a simple scheme 
based on a  quantum walk as illustrated in Fig.~\ref{fig:quantum_walk}, which 
can be implemented in a photonic system.
In a quantum walk \cite{Kurzy_ski_2013,Hou_2018,Tang_spin_2020,LiZhangZhu_2019}, 
the state $|x,c\>$ of
the walker-coin joint system is characterized by two indices $x$ and $c$, where 
$x=\dots,-1,0,1,\dots$ denotes the position of the walker on a one-dimensional chain, and $c=0,1$ labels the 
state of the coin qubit, which
determines the moving direction of the walker in the next step.
Each step of the quantum walk can be described by a unitary operator of the 
form $U(t)=\mathcal{T}C(t)$, where
$C(t)=\sum_x|x\>\<x|\otimes C(x,t)$, with $C(x,t)$ being
position-dependent coin operators, and $\mathcal{T}$ is the conditional translation operator,
\begin{equation}
	\mathcal{T} = \sum_x{(|x-1\rangle\langle x|\otimes|0\rangle\langle0| +
			|x+1\rangle\langle x|\otimes|1\rangle\langle1|)}.
\end{equation}

The ququart can be encoded into the initial state (with $t$ = 0) of the 
walker-coin system using the path degree of freedom (DoF). To be concrete, 
a general pure state $|\psi\rangle = \sum_{j=0}^{3}{a_j |j\rangle}$ of the 
ququart with $\sum_{j=0}^{3}{|a_j|^2} = 1$ is encoded as follows: 
\begin{equation}\label{eq:initialstate}
	|\Psi\rangle = (a_0|-3\rangle + a_1|-1\rangle + a_2|1\rangle + a_3|3\rangle) \otimes |1\rangle.
\end{equation}
After the first step of the quantum walk, the joint state evolves into
\begin{equation}
|\Psi_1\>=a_0|-2,1\>+a_1|-2,0\>+a_2|2,1\>+a_3|2,0\>.
\end{equation}
After the second step, the joint state evolves into
\begin{align}
|\Psi_2\>
=&\frac{\rmi a_0+a_1}{\sqrt{2}}|-3,0\>
+\frac{\rmi a_0-a_1}{\sqrt{2}}|-1,1\>\nonumber\\
&+\frac{a_2 + \rmi a_3}{\sqrt{2}}| 1,0\>
+\frac{a_2 - \rmi a_3}{\sqrt{2}}| 3,1\>.
\end{align}
Following a similar procedure, the final state after the quantum walk reads
\begin{align}
	|\Psi'\rangle
=&\frac{a_1 - \rmi a_0}{\sqrt{2}}|1,0\>
 -\frac{a_1 + \rmi a_0}{\sqrt{2}}|-1,1\>\nonumber\\
 &+\frac{a_2 - \rmi a_3}{\sqrt{2}}| 1,1\>
 +\frac{a_2 + \rmi a_3}{\sqrt{2}}|-1,0\>\nonumber\\
=&-\rmi\bigl(a_0|\Phi_0'\rangle + a_1|\Phi_1'\rangle
					+ a_2|\Phi_2'\rangle + a_3|\Phi_3'\rangle\bigr),
\end{align}
where 
\begin{equation}\label{eq:Phi'}
	\begin{aligned}
		|\Phi_0'\rangle &= \frac{1}{\sqrt{2}}(|+1,0\rangle + |-1,1\rangle),\\  
		|\Phi_1'\rangle &= \frac{\rmi}{\sqrt{2}}(|+1,0\rangle - |-1,1\rangle),\\ 
		|\Phi_2'\rangle &= \frac{\rmi}{\sqrt{2}}(|+1,1\rangle + |-1,0\rangle),\\ 
		|\Phi_3'\rangle &= \frac{1}{\sqrt{2}}(|+1,1\rangle - |-1,0\rangle). 
	\end{aligned}	
\end{equation}
Now $|\Psi'\rangle$ can be regarded as a two-qubit state on the Hilbert space 
$\mathcal{H}_A\otimes\mathcal{H}_B$, where $A$ denotes the effective walker 
qubit composed of positions $+1$ and $-1$, and $B$ denotes the coin qubit. Here 
$\{|\Phi_j'\rangle\}_{j=0}^{3}$ is exactly the two-qubit magic basis in 
$\mathcal{H}_A\otimes\mathcal{H}_B$, so 
the quantum walk illustrated in Fig.~\ref{fig:quantum_walk} indeed realizes 
the masker $M$ defined in \eref{eq:masker}. 

\emph{Experimental setup.}---The setup for masking the real ququart is shown 
in Fig.~\ref{fig:Experimental setup}. It contains 
four modules: a heralded single-photon source, a state-preparation module, a 
masking module, and a measurement 
module. Here we use the path DoF  to encode the real ququart 
and the polarization DoF  to encode the coin state employed in the quantum walk 
within the masking  module.

In the module of single-photon source, an ultraviolet laser with central 
wavelength of 404nm is used to pump a type-I phase-matched 
beta-barium borate (BBO) crystal to generate a photon pair in a product 
(polarization) state via spontaneous parametric 
down-conversion \cite{photon_source}. One photon is measured as a trigger to 
herald the generation of its twin photon, which is then transmitted to the 
state-preparation module. 

In the state-preparation module, a polarization beam 
splitter (PBS) first prepares the photon 
in the horizontal-polarization state $|H\rangle$ (in contrast with the 
vertical-polarization state $|V\rangle$). Then two beam displacers (BDs) 
and three half-wave plates (HWPs) transform the photon state into
\begin{equation}
 (a_0|-3\rangle + a_2|1\rangle) \otimes |H\rangle 
 + (a_1|-1\rangle + a_3|3\rangle) \otimes |V\rangle,   
\end{equation}
where the four real parameters $a_i$ for $i=0,...,3$ are controlled 
by three HWPs (see Appendix A). Finally, 
we prepare the ququart state 
\begin{equation}
    (a_0|-3\rangle + a_1|-1\rangle + a_2|1\rangle + a_3|3\rangle) \otimes |V\rangle
\end{equation}
of the form in \eref{eq:initialstate} by inserting a HWP at 45$^\circ$ 
(which realizes the  $X$ or NOT gate) on paths $-3$ and $1$ to turn $|H\rangle$ 
into $|V\rangle$. Note that the first two $X$ gates on paths $-1$ and $3$ in the 
masking module in Fig.~\ref{fig:quantum_walk} can  also be implemented by a HWP 
at 45$^\circ$; moreover, the  four paths can share a same HWP 
 placed at the beginning of the masking module in Fig.~\ref{fig:Experimental setup}.
The removable quarter-wave plate (QWP) in the preparation module is  used only to 
prepare pure states with complex coefficients (see Appendix~B).

After state preparation, the  ququart is sent into the  masking module, which 
realizes the quantum walk illustrated in Fig.~\ref{fig:quantum_walk}. The two 
QWP-HWP-QWP combinations realize the coin operators $C_1$ and $ C_2$, 
respectively, while the 
$0^\circ$ HWPs realize the $Z$ gates. The output of the masking module is a 
hybrid entangled state between the path qubit and
the polarization qubit of the heralded photon.

 The measurement module consists 
of two QWP-HWP combinations designed to control the measurement settings employed for measuring the path qubit and polarization qubit  (see Appendix C).  In addition, a K9 plate is used to compensate for different path lengths among 
the interference arms. Finally, the heralded photon is collected by four 
single-photon counting modules (SPCMs).

\begin{figure}[htbp]
	\center{\includegraphics[scale=0.146]{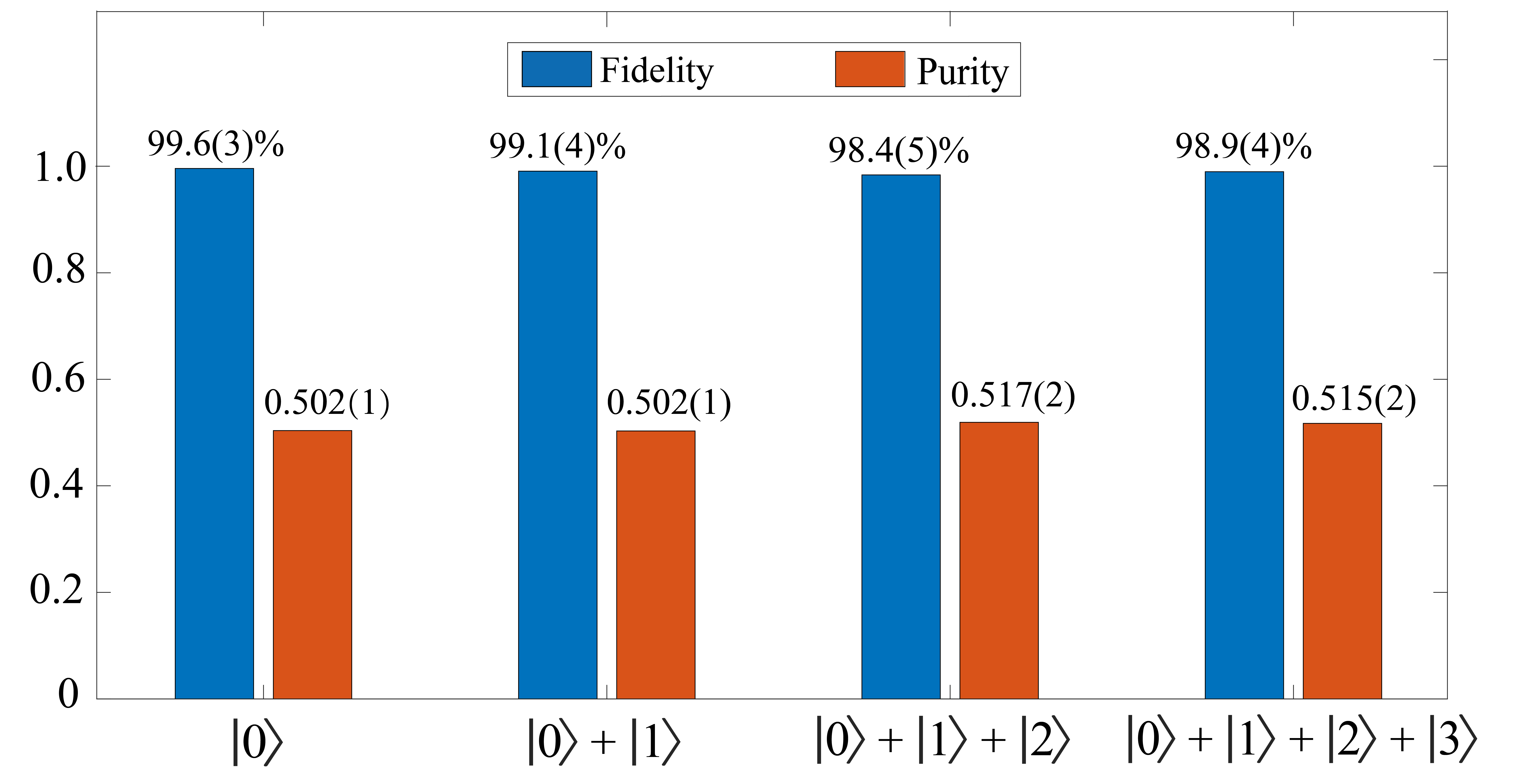}}
	\caption{\label{fig:fidelity_purity} Experimental characterization of 
	the masking module in Fig.~\ref{fig:Experimental setup}. Each blue bar 
	represents the fidelity between the actual output state and the ideal 
	output state associated with each probe state as marked at the bottom 
	of the figure. Each red bar represents the average purity of the two 
	reduced states of the output state. The number in the parentheses above the blue (red) bar
	indicates the $95\%$ confidence interval of each fidelity estimator (the standard deviation of each purity estimator). The  
	 methods for quantifying the confidence interval and the standard deviation are detailed 
	in Appendices D and E.
}
\end{figure}

\emph{Experimental results.}---To demonstrate the performance of the masking 
protocol realized using the photonic quantum walk as described above,  we select the 
following four probe states:  
\begin{equation}\label{eq:inputState}
\begin{gathered}
|0\rangle, \quad \frac{1}{\sqrt{2}}(|0\rangle + |1\rangle), \\
\frac{1}{\sqrt{3}}(|0\rangle + |1\rangle + |2\rangle), 
\quad \frac{1}{2}(|0\rangle + |1\rangle + |2\rangle + |3\rangle),
\end{gathered}
\end{equation}
which  form a complete but non-orthogonal basis in $\caH$. To verify that the 
masking module indeed transforms these input states according to the theoretical prediction described in 
Eqs.~(\ref{eq:initialstate})-(\ref{eq:Phi'}), we perform fidelity estimation 
based on quantum state verification (QSV) \cite{QSV, Zhu_2019PRL, Zhu_2019PRA} 
(see Appendix D) on the output states from 
the masking module. It turns out the fidelity  between each output state and the 
ideal masked state is  above 98\%, as shown in Fig.~\ref{fig:fidelity_purity}. 
To further certify that no information of the input state can be inferred from 
each reduced state of the output state, we then  perform quantum state 
tomography on  the path and polarization qubits, respectively. In each case the 
average  purity obtained in the experiment  is  very close to 0.5 as shown in 
Fig.~\ref{fig:fidelity_purity}, which implies that the reduced states of both 
subsystems are nearly  completely mixed and little information about the original 
state can be retrieved from each subsystem alone; see Appendix E for more  details. These experimental results clearly demonstrate that 
the masking module in Fig.~\ref{fig:Experimental setup} successfully realizes the 
masker $M$ in \eref{eq:masker} within small experimental errors.  

\begin{figure}[htbp]
	\center{\includegraphics[scale=0.31]{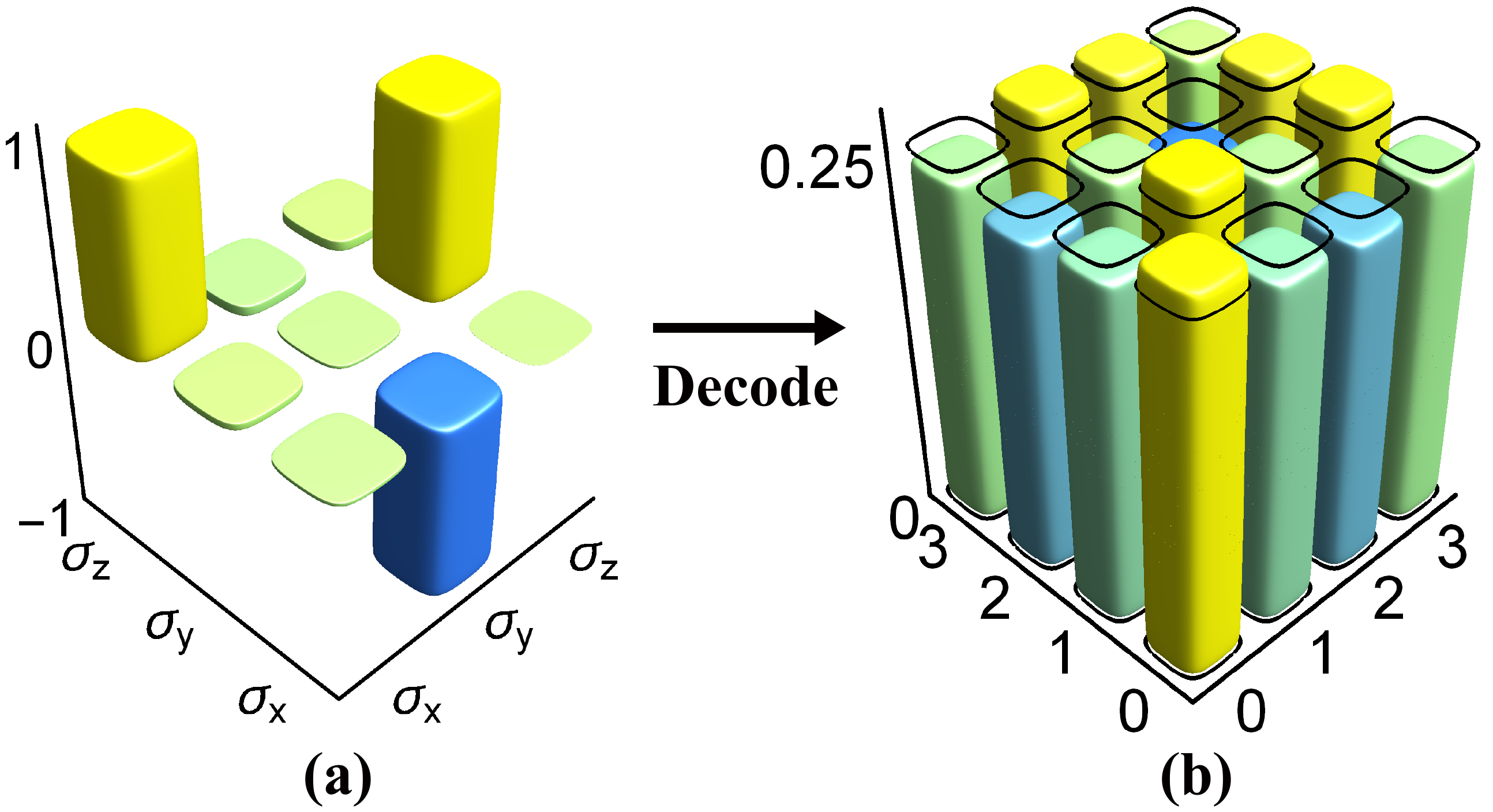}}
	\caption{\label{fig:tomo2}
	Experimental decoding  of the masked state from correlation measurements. Here the input state is given by $(|0\rangle + |1\rangle + |2\rangle + |3\rangle)/2$. (a) Mean values $\langle\sigma_i\otimes\sigma_j \rangle$ for $i, j \in \{x,y,z\}$. (b) Matrix elements of the reconstructed density matrix (only real part is shown since the imaginary part is zero by  reconstruction; see Appendix F). Bars without color represent the original input state, while bars with color represent the reconstructed state.}%
\end{figure}

Next, we show that the information of the input state can be faithfully 
retrieved from the bipartite correlations of the output state of the 
masker $M$. To be concrete, the density matrix of any real state can be 
reconstructed from the  correlation statistics of nine product Pauli 
measurements $\sigma_j\otimes \sigma_k$ for $j,k=x,y,z$ on the output state.  
In the experiment, each tensor product  $\sigma_j\otimes \sigma_k$ is measured 
4000 times to determine its mean value.
The specific decoding method is detailed in Appendix F. 
As an example, the decoding result on the input state 
$(|0\rangle + |1\rangle + |2\rangle + |3\rangle)/2$ is shown in 
Fig.~\ref{fig:tomo2}, and the decoding fidelity is 98.9\%, which 
is consistent with the result of QSV shown in Fig.~\ref{fig:fidelity_purity}.

\begin{figure}[htbp]
	\center{\includegraphics[scale=0.30]{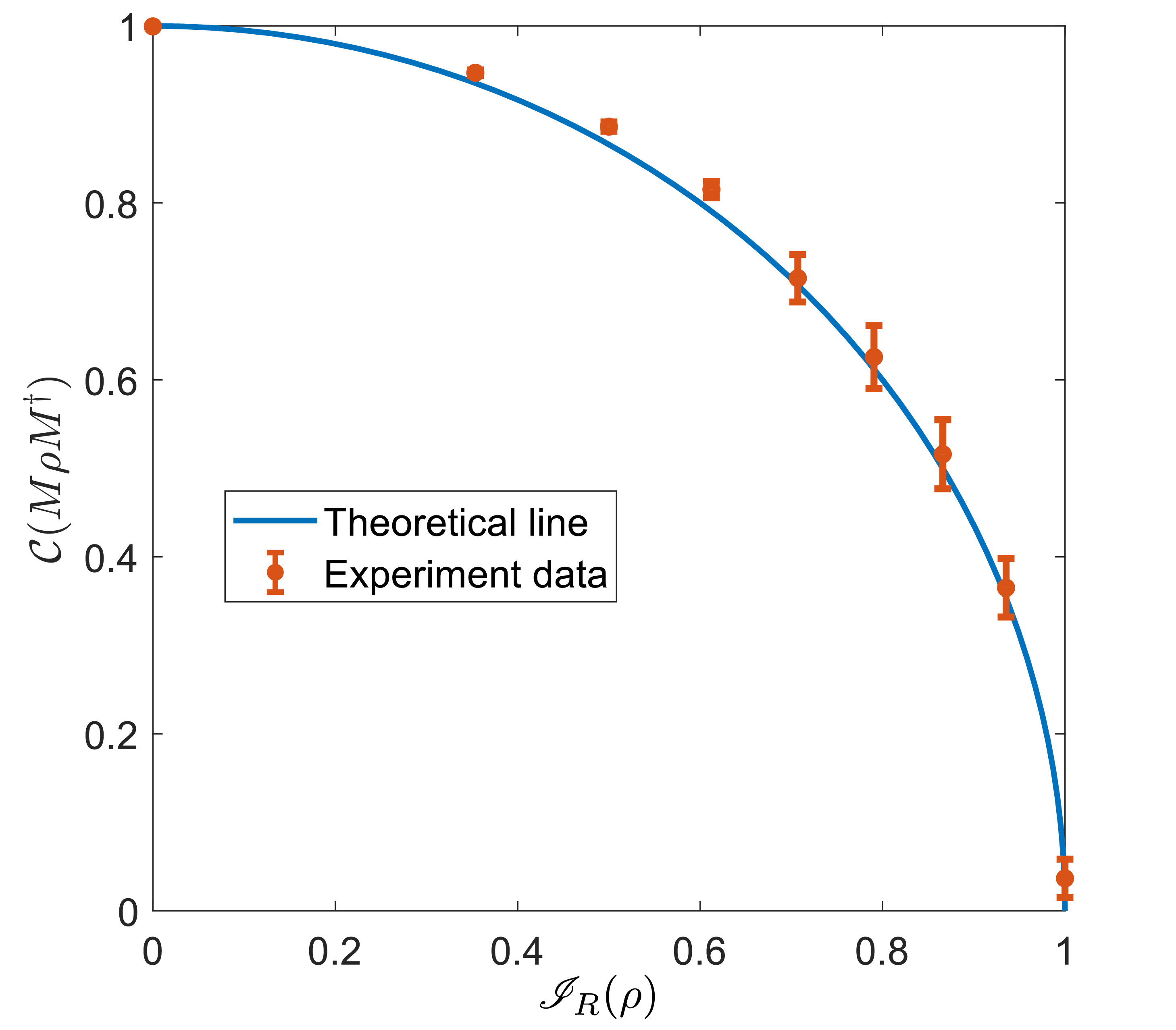}}
	\caption{\label{fig:concurrence}
Relation between the concurrence of the output state and the robustness 
of imaginarity of the input state, which has the form 
$(|0\rangle + e^{i\phi}|1\rangle)/{\sqrt{2}}$ with $0\leq \phi\leq \pi/2$. 
The error bars represent the standard deviations of the concurrence estimations, 
which are determined via re-sampling (see Appendix~E).}
\end{figure}

Finally, we verify the relation between the concurrence of the output state 
and the robustness of imaginarity of the input state as presented 
in \eref{eq:ConRI}. To be concrete, we choose  input states  of the form
\begin{equation}\label{eq:complex_state}
|\psi(\phi)\rangle =  \frac{1}{\sqrt{2}}(|0\rangle + e^{\rmi\phi}|1\rangle),\quad 0\leq \phi\leq \frac{\pi}{2}.
\end{equation}
The robustness of imaginarity of the input state and the concurrence of the output state are respectively given by 
\begin{equation}
\!\!\mathscr{I}_R(|\psi(\phi)\rangle) = \sin\phi,\quad  C(M|\psi(\phi)\rangle) = \cos\phi.
\end{equation}
In the experiment, the concurrence is determined by virtue of  the formula $\mathcal{C}(|\Psi\rangle) := \sqrt{2(1-\tr(\varrho_\rmP^2))}$, 
where $\varrho_\rmP$ is the reduced state of the path qubit of the output state as determined by 
quantum state tomography (see Appendix E). 
The experimental results shown in Fig.~\ref{fig:concurrence} 
agree very well with the theoretical prediction, which implies 
that  partial information of the 
input state is accessible to each subsystem once the robustness of 
imaginarity becomes nonzero.  These results also corroborate the 
conclusion that $\mathcal{D}^{\rmR}(\mathcal{H})$ is a maximal maskable set. 

\emph{Summary.}---We experimentally realized a masking protocol of the real 
ququart using a photonic quantum walk. This is the first experiment on quantum 
information masking beyond a qubit system.  Our  experiment clearly demonstrates 
that quantum  information of the real ququart can be completely hidden in 
bipartite correlations of hybrid entangled states,  which is not accessible 
to each subsystem alone, but can  be faithfully retrieved from  correlation 
measurements. By contrast, any superset of the set of real density matrices 
cannot be masked. Furthermore, the entanglement of the output state is tied to 
the robustness of imaginarity of the input state.  These results  manifest a 
sharp distinction between real quantum mechanics and complex quantum mechanics 
in hiding and masking quantum information. Moreover, they
offer valuable insights on the potential and  limitation of quantum information 
masking, which are of intrinsic interest to many active research areas.

It should be pointed out that 
 the two DoFs sharing 
the masked information in our experiment reside in the same location (photon), and  the meaning of "masking" is slightly different from the common literature. Here the information is masked into the correlations between two DoFs instead of  two space-separated particles.  Nevertheless, the underlying mathematical structures in the two scenarios are identical. So the main conclusions obtained in our proof-of-principle experiment should apply to both scenarios. Also, in principle our experiment can be generalized to realize quantum information masking in two photons by
cascading the protocol of quantum state fission 
\cite{quantum_fission} after the photonic quantum walk.

\section*{ACKNOWLEDGMENTS}
The work at the University of Science and Technology of China is supported by the National Natural Science Foundation of China (Grants Nos. 61905234, 11974335, 11574291, and 11774334), the Key Research Program of Frontier Sciences, CAS (Grant No. QYZDYSSW-SLH003) and the Fundamental Research Funds for the Central Universities (Grant No. WK2470000026).
The work at Fudan University is supported by  the National Natural Science Foundation of China (Grant No.~11875110) and  Shanghai Municipal Science and Technology Major Project (Grant No.~2019SHZDZX01).

\bigskip

\section*{APPENDIX A: Preparation of pure states with real coefficients}
\setcounter{equation}{0}
\setcounter{figure}{0}
\renewcommand{\theequation}{A\arabic{equation}}
\renewcommand{\thefigure}{A\arabic{figure}}
\begin{figure}[htbp]
	\center{\includegraphics[scale=0.6]{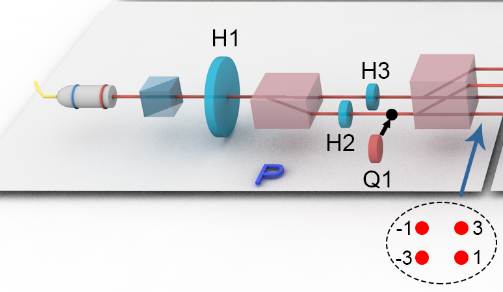}}
	\caption{\label{fig:preparation}
    The preparation module of the experimental setup.}
\end{figure}
In this section we show that the state-preparation module in Fig.~\ref{fig:Experimental setup}
 in the main text (also shown in Fig.~\ref{fig:preparation}) can prepare an arbitrary pure state of the real 
ququart. Suppose the angles of the optical axes of H1, H2, and H3 in Fig.~\ref{fig:preparation} are 
$h_1$, $h_2$, and $h_3$ (with respect to the horizontal direction), respectively. The input 
heralded photon is initially prepared in the state $|1\rangle\otimes|H\rangle$, which is turned 
into the following state
\begin{equation}
|1\rangle\otimes[\cos(2h_1)|H\rangle + \sin(2h_1)|V\rangle]
\end{equation}
by H1. The BD after H1 coherently routes the heralded photon to paths $-3$ and $1$
according to the polarization state of the heralded photon, which yields the state 
\begin{equation}
|-3\rangle\otimes\cos(2h_1)|H\rangle + |1\rangle\otimes\sin(2h_1)|V\rangle.
\end{equation}
Then H2 and H3 transform the photon state into
    \begin{align}
    &|-3\rangle\otimes\bigl[\cos(2h_1)\cos(2h_2)|H\rangle+\cos(2h_1)\sin(2h_2)|V\rangle\bigr] \nonumber\\
    &+ |1\rangle\otimes\bigl[\sin(2h_1)\sin(2h_3)|H\rangle-\sin(2h_1)\cos(2h_3)|V\rangle\bigr].
    \end{align}
Next, the second BD transforms the state of the heralded photon into
\begin{align}
    &\bigl[\cos(2h_1)\cos(2h_2)|-3\rangle + \sin(2h_1)\sin(2h_3)|1\rangle\bigr]\otimes|H\rangle \nonumber\\ +
    &\bigl[\cos(2h_1)\sin(2h_2)|-1\rangle - \sin(2h_1)\cos(2h_3)|3\rangle\bigr]\otimes|V\rangle.
\end{align}

Finally, by inserting $45^\circ$ HWPs on paths $-3$ and $1$, we can prepare the following state,
\begin{equation}
\begin{aligned}
    \bigl[&\cos(2h_1)\cos(2h_2)|-3\rangle + \cos(2h_1)\sin(2h_2)|-1\rangle\\ 
    &+ \sin(2h_1)\sin(2h_3)|1\rangle - \sin(2h_1)\cos(2h_3)|3\rangle\bigr]\otimes|V\rangle,
\end{aligned}
\end{equation}
which has  the form of \eref{eq:initialstate} in the main text. Incidentally, this operation can also be performed 
in the masking module, as pointed out in the main text. To prepare the state in \eref{eq:initialstate}, we need to 
choose the parameters $h_1$, $h_2$, and $h_3$ so as to satisfy the following equations:
\begin{equation}
    \begin{aligned}
        \cos(2h_1)\cos(2h_2) &= a_0, \\
        \cos(2h_1)\sin(2h_2) &= a_1, \\
        \sin(2h_1)\sin(2h_3) &= a_2, \\
        -\sin(2h_1)\cos(2h_3) &= a_3.
    \end{aligned}
\end{equation}

\section*{APPENDIX B: Preparation of pure states with complex coefficients}
In this section we show that the state-preparation module in Fig.~\ref{fig:Experimental setup}  in the main text (also shown in Fig.~\ref{fig:preparation}) can also prepare certain pure states with complex coefficients 
as presented in \eref{eq:complex_state} in the main text.
By inserting Q1 into path $-3$ as marked in Fig.~\ref{fig:preparation}, we can prepare the
states defined in \eref{eq:complex_state}. Such pure states with complex coefficients are 
required to demonstrate the relation between the concurrence of the output state and the 
robustness of imaginarity of the input state as presented in \eref{eq:ConRI} in the main text. 

To be specific, denote the angles of the optical axes of H1 and Q1 from the
horizontal direction by $h_1$ and $q_1$. Now we set $q_1 = 45^\circ$ and $h_1 = 0^\circ$, then the state of the heralded photon after Q1 reads \cite{Error}
\begin{align}
    &|-3\rangle\otimes\frac{1}{\sqrt{2}}\bigl\{[\cos(45^\circ - 2h_2) + \rmi \sin(45^\circ - 2h_2)]|H\rangle \nonumber \\
    &+ [\cos(45^\circ - 2h_2) - \rmi \sin(45^\circ - 2h_2)]|V\rangle\bigr\}.
\end{align}
If we set $h_2 = \frac{\phi}{4} + 22.5^\circ$ (here $\phi$ is the parameter appearing in \eref{eq:complex_state} in 
the main text), then the  photon state  after Q1 reduces to 
\begin{equation}
    |-3\rangle\otimes\frac{1}{\sqrt{2}}(|H\rangle + \rme^{\rmi\phi}|V\rangle)
\end{equation}
up to a global phase.
After the action of the second BD, the  photon state becomes
\begin{equation}
   \frac{1}{\sqrt{2}} (|-3\rangle|H\rangle + \rme^{\rmi\phi}|-1\rangle|V\rangle). 
\end{equation}
Finally, by inserting a 45$^\circ$  HWP  (which realizes the  $X$ or NOT gate) on paths $-3$ and $1$, we 
can prepare the following state
\begin{equation}
\frac{1}{\sqrt{2}}(|-3\rangle + \rme^{\rmi\phi}|-1\rangle)\otimes |V\rangle,
\end{equation}
which agrees with \eref{eq:complex_state} in the main text. Note that this HWP can also be considered as a part of 
the masking module.

\section*{APPENDIX C: Implementation of local projective measurements}

\begin{figure}[htbp]
	\center{\includegraphics[scale=0.35]{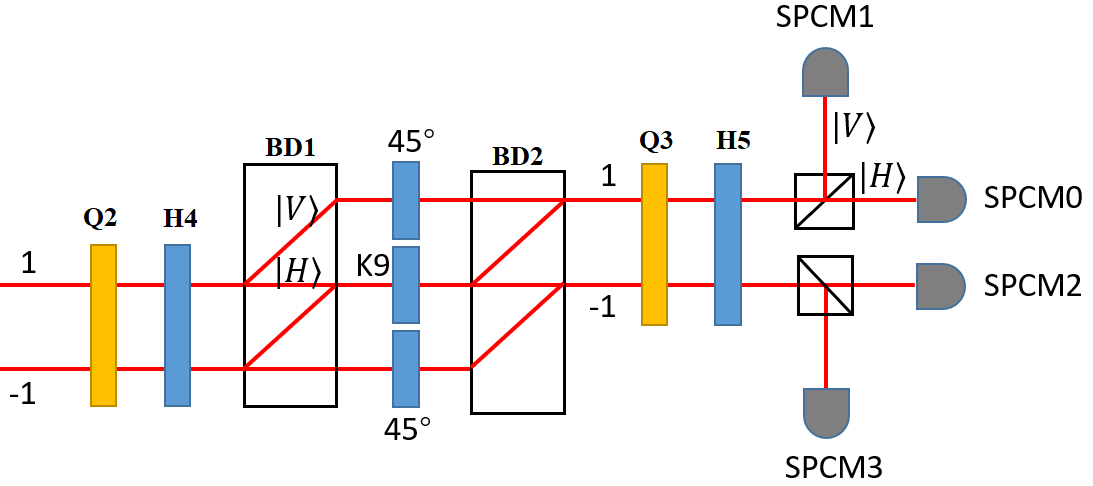}}
	\caption{\label{fig:measurement}
		The measurement module of the experimental setup.}
\end{figure}

In this section, we show that the measurement module in 
Fig.~\ref{fig:Experimental setup} (also shown in 
Fig.~\ref{fig:measurement}) can perform an arbitrary local 
projective measurement on the path-polarization two-qubit system. 
Suppose we want to perform the projective measurement onto the product basis
\begin{equation}\label{eq:ProdBasisPP}
\{|\varphi_0\rangle|\psi_0\rangle, |\varphi_0\rangle|\psi_1\rangle, |\varphi_1\rangle|\psi_0\rangle, |\varphi_1\rangle|\psi_1\rangle\},
\end{equation} 
 where 
\begin{equation}
    \begin{aligned}
         |\varphi_0\rangle &= \cos\gamma|0\rangle + \rme^{\rmi\zeta}\sin\gamma|1\rangle,\\
 |\varphi_1\rangle &= \sin\gamma|0\rangle - \rme^{\rmi\zeta}\cos\gamma|1\rangle,\\      
        |\psi_0\rangle &= \cos\alpha|0\rangle + \rme^{\rmi\beta}\sin\alpha|1\rangle,\\
        |\psi_1\rangle &= \sin\alpha|0\rangle - \rme^{\rmi\beta}\cos\alpha|1\rangle.
    \end{aligned}
\end{equation}
To simplify the notation, here we temporarily use $|0\rangle$ ($|1\rangle$) to refer to "path 1" 
("path $-1$") for the path qubit, and to "horizontal polarization" ("vertical polarization") for the 
polarization qubit. Then the output state from the masking module can be expressed as follows,
\begin{equation}
    |\Psi\rangle = a_0|\varphi_0\rangle|\psi_0\rangle + a_1|\varphi_0\rangle|\psi_1\rangle + 
                    a_2|\varphi_1\rangle|\psi_0\rangle + a_3|\varphi_1\rangle|\psi_1\rangle.
\end{equation}

We first choose the angles of the optical axes of Q2 and H4 in  Fig.~\ref{fig:measurement} so 
as to implement the transformation $|\psi_0\rangle \to |0\rangle,\ |\psi_1\rangle \to |1\rangle$ 
on the polarization qubit (see Supplemental Material of Ref.~\cite{Error} on how to calculate 
the angles), which turns $|\Psi\rangle$ into
\begin{equation}
    \begin{aligned}
        |\Psi'\rangle &= (a_0|\varphi_0\rangle + a_2|\varphi_1\rangle)\otimes|0\rangle 
        + (a_1|\varphi_0\rangle + a_3|\varphi_1\rangle)\otimes|1\rangle.
    \end{aligned}
\end{equation}
Then the two BDs split and re-combine the two light beams, which turns $|\Psi'\rangle$ into
\begin{equation}
    |\Psi''\rangle = |0\rangle \otimes (a_1|\varphi_0\rangle + a_3|\varphi_1\rangle) 
    + |1\rangle \otimes (a_0|\varphi_0\rangle + a_2|\varphi_1\rangle).
\end{equation}
Next, we set the angles of the optical axes of Q3 and H5 so as to realize the transformation 
$|\varphi_0\rangle \to |0\rangle, \ |\varphi_1\rangle \to |1\rangle$, which turns $|\Psi''\rangle$ into  
\begin{equation}
    |\Psi'''\rangle = a_1|0\rangle|0\rangle + a_3|0\rangle|1\rangle + a_0|1\rangle|0\rangle + a_2|1\rangle|1\rangle.
\end{equation} 
Now, the probabilities that the heralded photon is found by SPCM 0, 1, 2, 3 are 
$|a_1|^2$, $|a_3|^2$, $|a_0|^2$, $|a_2|^2$, respectively. In this way, we can realize 
the local projective measurement onto the product basis in \eref{eq:ProdBasisPP}.

\section*{APPENDIX D: Fidelity estimation based on quantum state verification}
To estimate the fidelity between  a two-qubit state $\varrho$ and the Bell state
$|\Phi\rangle=(|00\rangle + |11\rangle)/\sqrt{2}$, we can apply the idea of quantum 
state verification according to \rcite{QSV}. To be specific, we can randomly perform 
one of the three local projective tests associated with the three test projectors 
$P_{XX}^{+}$, $P_{YY}^{-}$, $P_{ZZ}^{+}$, respectively, where
\begin{equation}
\begin{aligned}
&P_{XX}^{+}=\frac{\mathbbm{1}+X\otimes X}{2}, \\ &P_{YY}^{-}=\frac{\mathbbm{1}-Y\otimes Y}{2}, \\ &P_{ZZ}^{+}=\frac{\mathbbm{1}+Z\otimes Z}{2}, 
\end{aligned}
\end{equation}
and $\mathbbm{1}$ stands for the identity operator acting on the two-qubit system. To 
optimize the performance, each test is performed with probability $1/3$. The resulting 
verification operator reads
\begin{equation}
    \Omega = \frac{1}{3}(P_{XX}^{+}+P_{YY}^{-}+P_{ZZ}^{+}) = |\Phi\rangle\langle\Phi| + \frac{1}{3}(\mathbbm{1}-|\Phi\rangle\langle\Phi|).
\end{equation}

Now suppose that the fidelity between $\varrho$ and $|\Phi\rangle\langle\Phi|$ 
is $F(\varrho, |\Phi\rangle)= \langle\Phi|\varrho|\Phi\rangle = 1 - \epsilon$, 
where $\epsilon$ is the infidelity. Then the probability that 
$\varrho$ can pass each test on average is given by
\begin{equation}
    p_{\mathrm{succ}} = {\rm tr}(\varrho\Omega) = 1-\frac{2}{3}\epsilon,
\end{equation}
which implies that
\begin{equation}
\epsilon = \frac{3}{2}(1 - p_{\mathrm{succ}}).
\end{equation}
If $\varrho$ passes $S$ tests after $N$ tests in total in a given experiment, then 
$\hat{p}_{\mathrm{succ}}:=S/N$ is an unbiased estimator for $p_{\mathrm{succ}}$, 
from which we can construct an unbiased estimator for the infidelity,
\begin{equation}
\hat{\epsilon} = \frac{3}{2}(1 - \hat{p}_{\mathrm{succ}}).
\end{equation}

To calculate the 95\% confidence interval of $\hat{\epsilon}$, here we 
assume that all the states $\varrho$  in the $N$ runs are independent and identically 
distributed. A common choice for  the 95\% confidence interval of 
$\hat{p}_{\mathrm{succ}}$ is 
the Agresti-Coull interval \cite{binomial_interval}, which has the form
\begin{equation}
    \tilde{p} - \kappa_{\alpha}(\tilde{p}\tilde{q})^{1/2}\tilde{N}^{-1/2} \le p_{\mathrm{succ}} \le \tilde{p} + \kappa_{\alpha}(\tilde{p}\tilde{q})^{1/2}\tilde{N}^{-1/2},
\end{equation}
where 
\begin{equation}
\tilde{p} = \frac{\tilde{S}}{\tilde{N}},\quad  \tilde{q} = 1 - \tilde{p}, \quad \tilde{S} = S + \frac{\kappa_{\alpha}^2}{2},\quad \tilde{N} = N + \kappa_{\alpha}^2,
\end{equation}
and $\kappa_{\alpha} := z_{\alpha/2}$ is the ($1 - \alpha/2$)th quantile of the standard normal distribution. Accordingly, the 95\% confidence interval of $\hat{\epsilon}$ reads
\begin{equation}\label{eq:binomial_interval}
    \epsilon_{\rm low} \le \epsilon \le \epsilon_{\rm high},
\end{equation}
where 
\begin{equation}
\begin{aligned}
    &\epsilon_{\rm low} = \frac{3}{2}\bigl[1 - \tilde{p} - \kappa_{0.05}(\tilde{p}\tilde{q})^{1/2}\tilde{N}^{-1/2}\bigr],\\ 
    &\epsilon_{\rm high} = \frac{3}{2}\bigl[1 - \tilde{p} + \kappa_{0.05}(\tilde{p}\tilde{q})^{1/2}\tilde{N}^{-1/2}\bigr].
\end{aligned}
\end{equation}

When the target state changes from $|\Phi\rangle$ to, say, 
$|\Phi'\rangle = (U\otimes \mathbbm{1}_B)|\Phi\rangle$, the above approach still 
applies as long as the set of test projectors is replaced by 
$\{P_{X'X}^{+}, P_{-Y'Y}^{+}, P_{Z'Z}^{+}\}$ accordingly, where $X' = UXU^{\dagger}$, $Y' = UYU^{\dagger}$, 
$Z' = UZU^{\dagger}$. 

In our experiment, the total number of tests is chosen to be $N = 5000$ to reduce the 
statistical fluctuation.  The values of $S$ associated with the four probe states 
indicated in Fig.~3 in the main text are $4986$, $4969$, $4946$, and $4963$, respectively. 
The estimation errors of the fidelities shown in Fig.~3  are determined 
by virtue of these numbers and the formula
\begin{align}
\max\{\hat{\epsilon}-\epsilon_{\rm low}, \epsilon_{\rm high}-\hat{\epsilon}\}.
\end{align}
Note that the estimation error of each fidelity is the same as the estimation error of each infidelity.


\section*{APPENDIX E: Quantum state tomography of the path qubit and polarization qubit}

\begin{figure*}[htbp]
	\center{\includegraphics[scale=0.6]{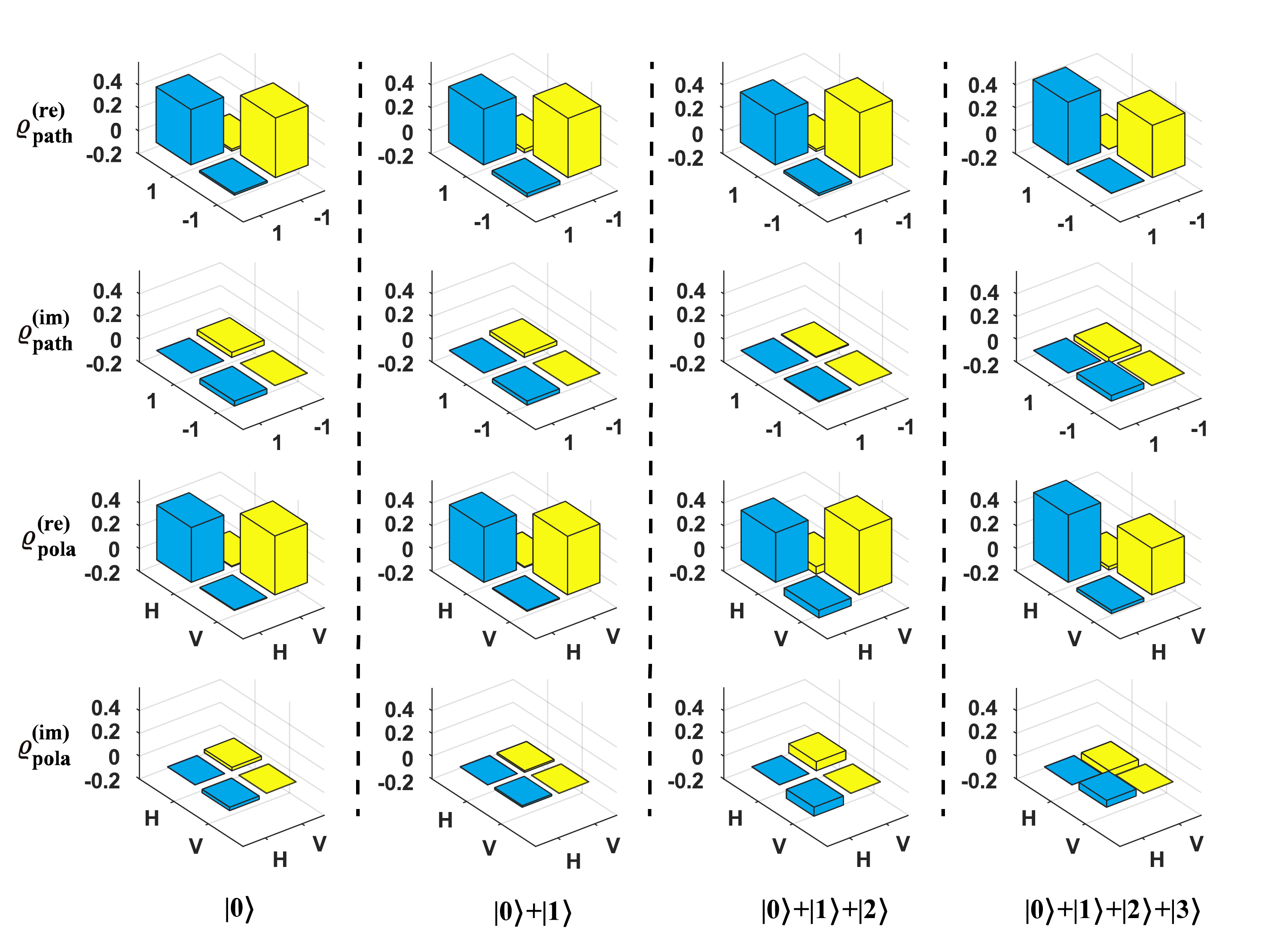}}
	\caption{\label{fig:tomo1}
		Experimental results on the reduced density matrices of the path qubit ($\varrho_{\rm path}$) and 
		polarization qubit ($\varrho_{\rm pola}$), respectively, of the output states of the masker defined in \eref{eq:masker} in the main text. 
		Here "re" denotes the real part, and "im" denotes the imaginary part. The bottom line marks the 
		input probe states associated with the output states.
		The experimental results demonstrate that all the reduced density matrices are nearly maximally mixed.}
\end{figure*}

To further characterize the performance of the masking module, we perform quantum state 
tomography on the path qubit and polarization qubit of the output states associated with 
the four input states $|0\rangle$, $(|0\rangle + |1\rangle)/\sqrt{2}$, 
$(|0\rangle + |1\rangle + |2\rangle)/\sqrt{3}$, and $(|0\rangle + |1\rangle + |2\rangle + |3\rangle)/2$,  
respectively [cf. Eq.~(13) in the main text]. To determine each reduced density matrix of each 
output state, $X$, $Y$, $Z$ measurements are performed 4000 times, respectively. Then the density 
matrix is reconstructed using the maximum-likelihood method described in \rcite{tomo}. The reconstruction results on 
the reduced density matrices are shown in Fig.~\ref{fig:tomo1}.

By virtue of the above results we can compute the average purity of the two reduced density matrices 
of each output state, as shown in Fig.~3 in the main text. To determine the estimation error of the 
average purity,
we re-sample $100$ times the photon-counting  results used in the  tomography according to the 
Poisson distribution, and then calculate the standard deviation of the average purity using the 
re-sampled data.

Since the output state is pure, its concurrence is determined by the purity of each reduced density 
matrix, which in turn can be determined by quantum state tomography. The concurrence shown in Fig.~5 
in the main text is inferred from the purity of the density matrix of the  path qubit using a similar 
way as mentioned above except that here $X$, $Y$, $Z$ measurements are performed 10000 times, respectively. 
To determine the estimation error of the concurrence,
we re-sample $100$ times the photon-counting  results used in the  tomography of the path qubit according to the 
Poisson distribution, and then calculate the standard deviation of the concurrence using the re-sampled data.




\begin{widetext}

\section*{APPENDIX F: Decoding input ququart from correlation measurements}
In the main text, we introduced  a simple masker $M:\caH\mapsto\caH_A\otimes\caH_B$ for the real 
ququart associated with the Hilbert space $\caH$, where both $\caH_A$ and $\caH_B$ are qubit Hilbert spaces. To be specific, 
 $M$ is the isometry defined by its action on the computational basis: 
\begin{equation}
M|j\> = -\rmi(U_j\otimes\mathbbm{1}_B)|\Phi\>, \qquad j=0,1,2,3,  
\end{equation}
where $U_0:=\mathbbm{1}_A$,  $\{U_j\}_{j=1}^3:=\{\rmi\sigma_z, \rmi\sigma_x, \rmi\sigma_y\}$ forms a 
set of Hurwitz-Radon matrices on $\caH_A$, and $|\Phi\>=(|00\>+|11\>)/\sqrt{2}$ is the canonical maximally entangled state in $\caH_A\otimes\caH_B$. 
Here we shall show that the information of the real input state
can be faithfully decoded from the bipartite correlations of the output state of the masker $M$.

Recall that any two-qubit state $\varrho$ on $\caH_A\otimes\caH_B$ can be expressed as
\begin{equation}
    \varrho = \frac{1}{4}\bigg(\mathbbm{1} + \bm{a}\cdot{\boldsymbol \sigma} \otimes \mathbbm{1}_B
            + \mathbbm{1}_A\otimes\bm{b}\cdot{\boldsymbol \sigma} 
            + \sum_{j,k}{T_{jk}}\,\sigma_j\otimes\sigma_k\bigg),
\end{equation}
where $\bm{\sigma}=(\sigma_x,\sigma_y,\sigma_z)$, $\bm{a}$ and $\bm{b}$ are the Bloch vectors of the two reduced states, respectively, and $T$ is the correlation matrix.
The entries of the correlation matrix can be determined by suitable
Pauli measurements according to the following equation
\begin{equation}
    T_{jk} =\langle \sigma_j\otimes\sigma_k \rangle= 
    {\rm tr}[\varrho(\sigma_j\otimes\sigma_k)], \quad j, k \in \{x, y, z\}.
\end{equation}

To decode the input state, we first consider the case in which the input real state is pure. 
A general real pure state in $\caH$ can be written as $|\psi(\bm{c})\>=\sum_jc_j|j\>$, 
where $\bm{c}=(c_0,c_1,c_2,c_3)$ is a normalized real vector. 
The corresponding output state of the masker $M$ has the form 
\begin{equation}
    |\Psi(\bm{c})\> := M|\psi(\bm{c})\> = -\rmi[U(\bm{c})\otimes\mathbbm{1}_B]|\Phi\>,
\end{equation}
where $U(\bm{c}) := \sum_{j=0}^3 {c_jU_j}$. Note that
\begin{equation}\label{eq:Phi}
    |\Phi\rangle\langle\Phi| = \frac{1}{4}(\mathbbm{1} + \sigma_x\otimes\sigma_x 
    - \sigma_y\otimes\sigma_y + \sigma_z\otimes\sigma_z),
\end{equation}
and
\begin{equation}
    \begin{aligned}
    U(\bm{c})\sigma_x U(\bm{c})^{\dagger} &= (c_0^2 - c_1^2 + c_2^2 - c_3^2)\sigma_x +
                                               2(c_2c_3-c_0c_1)\sigma_y + 2(c_0c_3+c_1c_2)\sigma_z,\\
    U(\bm{c})\sigma_y U(\bm{c})^{\dagger} &= 2(c_0c_1+c_2c_3)\sigma_x + 
                                                   (c_0^2 - c_1^2 - c_2^2 + c_3^2)\sigma_y + 2(c_1c_3-c_0c_2)\sigma_z,\\
    U(\bm{c})\sigma_z U(\bm{c})^{\dagger} &=  2(c_1c_2-c_0c_3)\sigma_x + 2(c_1c_3+c_0c_2)\sigma_y 
                                                  +(c_0^2 + c_1^2 - c_2^2 - c_3^2)\sigma_z.
    \end{aligned}
\end{equation}
So the density operator of $|\Psi(\bm{c})\>$ reads
\begin{equation}
    |\Psi(\bm{c})\>\<\Psi(\bm{c})| = \frac{1}{4}\bigg[\mathbbm{1} 
    + \sum_{j,k}{T(\bm{c})_{jk}}\,\sigma_j\otimes\sigma_k\bigg], 
\end{equation}
where the correlation matrix $T(\bm{c})$ has the form
\begin{equation}\label{eq:correlationT}
 T(\bm{c}) = 
 \left(
 \begin{matrix}
   c_0^2 - c_1^2 + c_2^2 - c_3^2 & -2(c_0c_1+c_2c_3) & 2(c_1c_2-c_0c_3) \\
   2(c_2c_3-c_0c_1) & -(c_0^2 - c_1^2 - c_2^2 + c_3^2) & 2(c_1c_3+c_0c_2) \\
   2(c_0c_3+c_1c_2) & -2(c_1c_3-c_0c_2) & c_0^2 + c_1^2 - c_2^2 - c_3^2
  \end{matrix}
  \right).
\end{equation}
In conjunction with the normalization condition $c_0^2 + c_1^2 + c_2^2 + c_3^2 = 1$,
we can deduce that
\begin{equation}\label{eq:cj^2}
    \begin{aligned}
        &c_0^2 = \frac{1}{4}(1 + T_{11} - T_{22} + T_{33}),\qquad
        &c_1^2 = \frac{1}{4}(1 - T_{11} + T_{22} + T_{33}),\\
        &c_2^2 = \frac{1}{4}(1 + T_{11} + T_{22} - T_{33}),\qquad
        &c_3^2 = \frac{1}{4}(1 - T_{11} - T_{22} - T_{33}),
    \end{aligned}
\end{equation}
and
\begin{equation}\label{eq:cjk}
    \begin{aligned}
        &c_2c_3 = \frac{1}{4}(T_{21} - T_{12}),\qquad
        &c_1c_3 = \frac{1}{4}(T_{23} - T_{32}),\qquad
        &c_1c_2 = \frac{1}{4}(T_{13} + T_{31}),\\ 
        &c_0c_1 = -\frac{1}{4}(T_{21} + T_{12}),\qquad
        &c_0c_2 = \frac{1}{4}(T_{23} + T_{32}),\qquad
        &c_0c_3 = \frac{1}{4}(T_{31} - T_{13}).
    \end{aligned}
\end{equation}

By linearity the above result can be generalized to any mixed 
input state $\rho$ that has a real density matrix. 
In this case,  in analogy to \eref{eq:correlationT}, the correlation matrix of the output 
state $M\rho M^\dag$ reads
\begin{equation}
    T = \left(
    \begin{matrix}
      \rho_{00} - \rho_{11} + \rho_{22} - \rho_{33} & -2(\rho_{01}+\rho_{23}) & 2(\rho_{12}-\rho_{03}) \\
      2(\rho_{23}-\rho_{01}) & -(\rho_{00} - \rho_{11} - \rho_{22} + \rho_{33}) & 2(\rho_{13}+\rho_{02}) \\
      2(\rho_{03}+\rho_{12}) & -2(\rho_{13}-\rho_{02}) & \rho_{00} + \rho_{11} - \rho_{22} - \rho_{33}
     \end{matrix}
     \right).
\end{equation}
Accordingly, we have
\begin{equation}
    \begin{aligned}
        &\rho_{00} = \frac{1}{4}(1 + T_{11} - T_{22} + T_{33}),\qquad
        &\rho_{11} = \frac{1}{4}(1 - T_{11} + T_{22} + T_{33}),\\
        &\rho_{22} = \frac{1}{4}(1 + T_{11} + T_{22} - T_{33}),\qquad
        &\rho_{33} = \frac{1}{4}(1 - T_{11} - T_{22} - T_{33}),
    \end{aligned}
\end{equation}
and
\begin{equation}
    \begin{aligned}
&\rho_{23} = \frac{1}{4}(T_{21} - T_{12}),\qquad
&\rho_{13} = \frac{1}{4}(T_{23} - T_{32}),\qquad
&\rho_{12} = \frac{1}{4}(T_{13} + T_{31}),\\
&\rho_{01} = -\frac{1}{4}(T_{21} + T_{12}),\qquad
&\rho_{02} = \frac{1}{4}(T_{23} + T_{32}),\qquad
&\rho_{03} = \frac{1}{4}(T_{31} - T_{13}),
    \end{aligned}
\end{equation}
which generalize \esref{eq:cj^2} and \eref{eq:cjk}.
Therefore, the density matrix of any real input state of the  ququart can be faithfully 
decoded from the correlation matrix of the output state, 
which can  easily be determined by virtue of  product Pauli measurements $\sigma_j\otimes \sigma_k$ for 
$j,k\in\{x,y,z\}$. 
\end{widetext}

\end{document}